\documentclass{aa}  

\def \half {\textstyle{\frac{1}{2}}}

\def \. {\cdot}

\renewcommand{\vec}[1]{{\mathbf #1}}

\newcommand{\uvec}[1]{ \hat{\mathbf #1} }

\newcommand{\xv}{ \uvec x }
\newcommand{\yv}{ \uvec y }
\newcommand{\zv}{ \uvec z }

\newcommand{\rone}{ r_1}
\newcommand{\rtwo}{ r_2}

\newcommand{\alfven}{Alfv{\'e}n\,}

\usepackage{graphicx}

\usepackage{txfonts}
\begin{document}

   \title{Chromospheric and coronal heating and jet acceleration due to reconnection driven by flux cancellation}

   \subtitle{II. Cancellation of two magnetic polarities of unequal flux}
   \titlerunning{Heating by Flux Cancellation}

   \author{P. Syntelis
          \inst{1}
          \and
          E. R. Priest\inst{1}
          }

   \institute{School of Mathematics and Statistics, University of St. Andrews, Fife KY16 9SS, Scotland, UK\\
              \email{psyntelis@gmail.com}
             }

   \date{Received ; accepted }

% \abstract{}{}{}{}{} 
% 5 {} token are mandatory
 
  \abstract
  % {} leave it empty if necessary  v
   {Recent observations have shown that magnetic flux cancellation occurs at the photosphere more frequently than previously thought.}
   {In order to understand the energy release by reconnection driven by flux cancellation, we previously studied a simple model of two cancelling polarities of equal flux. Here, we further develop  our analysis to achieve a more general setup where the two cancelling polarities have unequal magnetic fluxes and where many new features are revealed. }
   {We carried out an analytical study  of the cancellation of two magnetic fragments of unequal and opposite flux that approach one another and are located in an overlying horizontal magnetic field.}
   {The energy release as microflares and nanoflares occurs in two main phases. During phase\ 1a, a separator is formed and reconnection is driven at it as it rises to a maximum height and then moves back down to the photosphere, heating the plasma and accelerating plasma jets in the process. During phase 1b, once the separator moves back to the photosphere, it bifurcates into two null points. Reconnection is no longer driven at the separator and an isolated magnetic domain connecting the two polarities is formed. 
   During phase 2, the polarities cancel out at the photosphere as magnetic flux submerges below the photosphere and as reconnection occurs at and above the photosphere and plasma jets and a mini-filament eruption can be produced. }
   {}


  \maketitle
%
%-------------------------------------------------------------------

\section{Introduction} 
\label{sec:introduction}

Magnetic flux cancellation is the process whereby two opposite polarities approach each other, interact via magnetic reconnection, and eventually submerge into the solar interior \citep[e.g.][]{harvey85,martin85,priest87, vanBallegooijen_etal1989}. 
In recent years, flux cancellation has been associated with a variety of solar phenomena and its importance has been re-examined.
For example, observations above cancellation regions have revealed that flux cancellation is commonly associated with localised energy release events, such as Ellerman and IRIS bombs, UV and EUV bursts, and the corresponding reconnection-driven outflows and jets above such regions \citep[e.g.][]{
Watanabe_etal2011,
Vissers_etal2013,
Peter_etal2014,Rezaei_2015,Tian_etal2016,Reid_etal2016,Hong_etal2017,RouppevanderVoort2017, Ulyanov_2019, Huang_etal2019, Chen_etal2019,Ortiz_etal2020,Park_2020}.
Another example is eruption-driven jets, where flux cancellation is often associated with the formation and triggering of erupting mini-filaments  \citep[e.g.][]{sterling15,Panesar_etal2016,sterling16b,Panesar_etal2019}.

Aside from bursts and jets, observations have suggested that the cancellation of opposite polarities at the footpoints of coronal loops could be responsible for the brightening of coronal loops \citep[][]{tiwari14,chitta17b,huang18,chitta18,Sahin_etal2019}. 
Potentially, even spicules could be linked to flux cancellation \citep{sterling16a, Samanta_etal2019}.
Recent observations from the Sunrise balloon mission \citep{solanki10, solanki17}, with a spatial resolution six times better than that of the Helioseismic Imager (HMI) on the Solar Dynamics Observatory (SDO), have shown that flux cancellation is occurring at  rate that is an order of magnitude higher than previously thought \citep{smitha17}.

Inspired by these observations, \cite*{Priest_etal2018}  proposed that nanoflares driven by flux cancellation could heat the solar chromosphere and corona. Their model was further developed with numerical simulations, demonstrating the validity of the theoretical model and showing that a variety of reconnection-driven jets with different properties can be formed by flux cancellation \citep{Syntelis_etal2019, Syntelis_Priest_2020}.

This first presentation of this flux-cancellation model described the reconnection, examined above, which is driven by the convergence of two approaching polarities of equal flux in the presence of a horizontal ambient field. The analysis treated the current sheet as a  two-dimensional object. In a new series of papers, we are further developing  the model by treating the current sheet as an inherently three-dimensional structure and by extending the model to other geometries. 
As a first step, \citet{Priest_Syntelis2021} (herafter paper I) derived the magnetic field strength in the vicinity of a three-dimensional toroidal current, taking  its curvature into account . They also worked out the  resulting three-dimensional correction to  the model of \cite*{Priest_etal2018}.

\begin{figure*}[h!]
    \centering
    \includegraphics[width=0.7\textwidth]{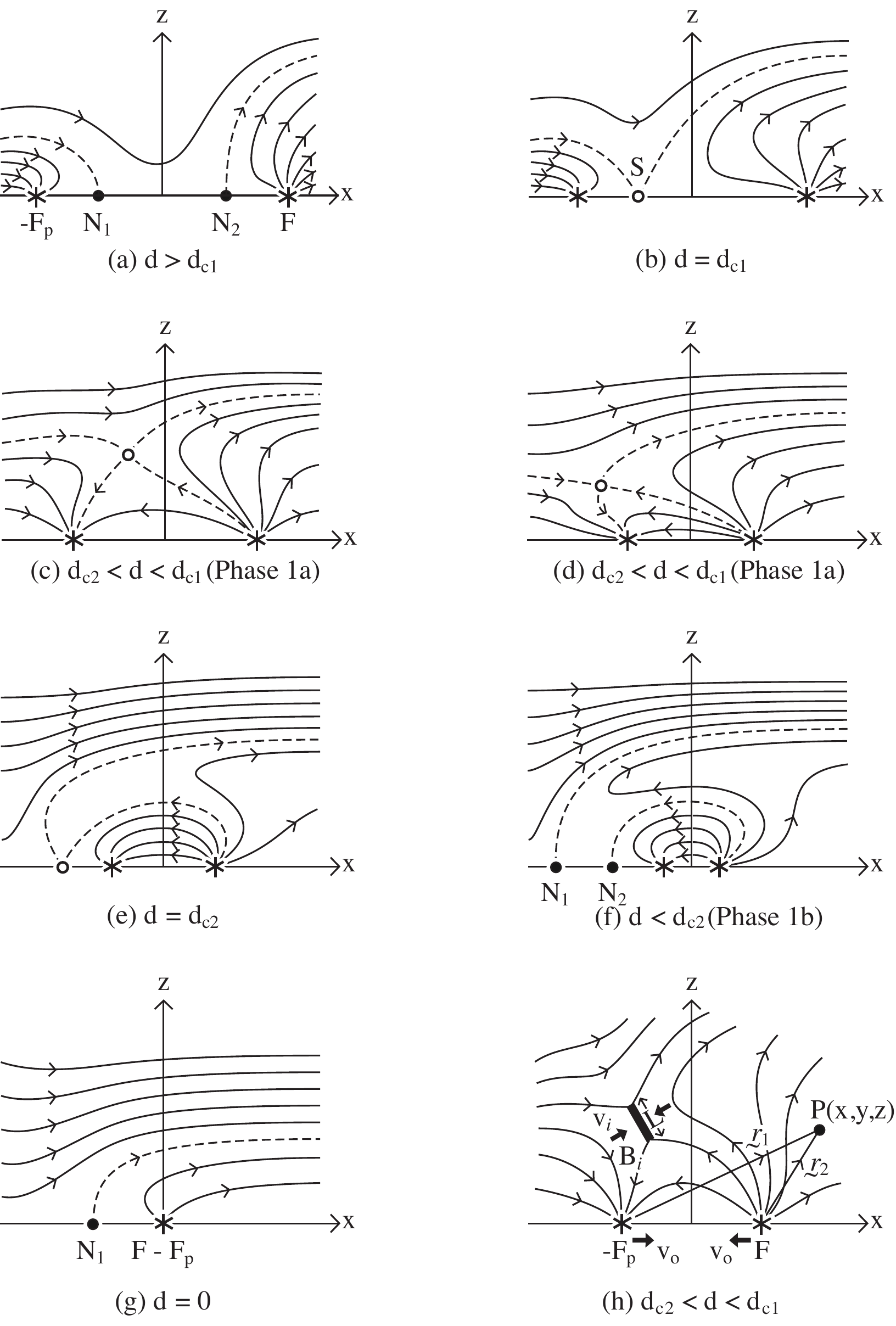}
    \caption{Overall behaviour of the magnetic field during the approach of two oppositely directed photospheric polarities (stars) of flux, -$F_p$ and $F,$ (where $|F_p|<|F|$) are separated by a distance, 2d, and situated in an overlying uniform horizontal magnetic field ($B_0$).
    The magnetic field is axisymmetric, shown here in a vertical ($y=0$) plane.  Dashed curves show separatrix magnetic field lines and solid curves indicate other  magnetic field lines. 
    The separator is marked by an unfilled dot, and null points with  solid dots. The field is shown for
        (a)  $d > d_{c_1}$, when the two sources are far apart, 
        (b)  $d = d_{c_1}$, when the nulls coalesce at the photosphere ($z = 0$) and a separator first appears, 
        (c)  $ d_{c_2} < d < d_{c_1}$ (phase 1a), when the separator arches above the surface,
        (d)  $ d_{c_2} < d < d_{c_1}$ (phase 1a), when the separator has moved to the left of the flux source $-F_p$,
        (e)  $d = d_{c_2}$, when the separator falls back down to the photosphere and becomes a second-order null,
        (f)  $0<d<d_{c_2}$ (phase 1b), when the null has bifurcated into two null points moving away from each other, and
        (g)  $d=0$, when the two sources have coalesced, leaving behind the remaining magnetic fragment of flux $(F-F_p)$.
        Panel (h) shows the notation used to describe the reconnection process at a current sheet of length $L$.} 
    \label{fig:cartoon}
\end{figure*}
In the present work (Paper II), we take the next step in the development of the theory by examining more realistic magnetic  configurations, where the two cancelling polarities have different fluxes. Thus, in this study, we examine the reconnection between two cancelling opposite polarities of unequal flux in the presence of a horizontal field. 
In Section~\ref{sec:theory}, we present the theoretical model. In Sections~\ref{sec:discussion} and \ref{sec:conclusions}, we discuss our results and conclusions.

\section{Theory of energy release at a
reconnecting current sheet driven by the cancellation of unequal polarities}
\label{sec:theory}

\subsection{Magnetic configuration}
Let us consider two sources of flux in the photospheric $xy$-plane. Source 1 has positive flux  ($F$) and is situated at $(d,0)$, whereas source 2 is a parasitic polarity with negative flux  ($-F_p$) with $F_p \leq $ F, and is situated at $(-d,0)$. The local overlying magnetic field ($B_0 \mathbf{\hat{x}}$) is horizontal and is aligned with the two sources.  The two sources approach one another, moving along the line that joins them at velocities that are slow enough ( $\pm v_0$) so that the field passes through a series of equilibria.

The overall behaviour of the magnetic field during the approach of the sources is shown in Figure~\ref{fig:cartoon}, which depicts a vertical section through an axisymmetric configuration. 
It summarises schematically the slow-time evolution of the system through a series of equilibrium states.
When the two sources, indicated by stars, are so far apart that they are not connected magnetically,  their fluxes are bound by axisymmetric separatrix surfaces, shown as dashed lines that meet the $x$-axis at 3D null points N$_1$ and N$_2$ (Fig.\ref{fig:cartoon}a). 

As the sources approach one another, the nulls eventually coalesce to form a second-order null, which then becomes a separator (S) that rises above the photosphere (Fig.\ref{fig:cartoon}c). Reconnection is driven at S, which consists of an axisymmetric ring of null points and about which a separatrix current sheet forms (Fig.\ref{fig:cartoon}h). We note that a slight perturbation of the axisymmetric configuration would turn S into a separator having a magnetic field component along it and join two 3D nulls that lie in the photosphere. 
Eventually, the separator moves back down to the photosphere, where it becomes a null point that splits into two nulls (Fig.\ref{fig:cartoon}f) and which themselves eventually coalesce as the flux sources combine to leave behind a source of magnitude $(F-F_p)$ (Fig.\ref{fig:cartoon}g). We now proceed to work out some of the details of this process.

The magnetic field above the photosphere ($z>0$) is then given by:
\begin{equation}
    \vec B = \frac{F}{2\pi} \frac{\vec\rone} {\rone^3} - \frac{F_p}{2\pi} \frac{\vec\rtwo}{\rtwo^3} + {B_0}\xv
\label{eq1}
,\end{equation}
where 
\begin{equation}
\vec \rone = (x+d) \xv + y \yv + z \zv, \ \ \ \ \ \ \ 
\vec \rtwo = (x-d) \xv + y \yv + z \zv \nonumber
\end{equation}
are the position vectors of a point $P(x,y,z)$ with respect to the two sources.

Magnetic fields are non-dimensionalised with respect to $B_0$,  fluxes with respect to $F$ and lengths with respect to the interaction distance \citep{Longcope_1998}:
\begin{equation}
    d_0 = \left( \frac{F}{\pi B_0} \right)^{1/2}.
\end{equation}
The resulting dimensionless variables are
$\bar{B}={B}/{B_{0}}$, 
$\bar{d} ={d}/{d_{0}}$, 
$\bar{r}={r}/{d_{0}},$ and
$\bar{F}_p={F_p}/{F},$
while Eq. (\ref{eq1}) becomes:
\begin{equation}
    \bar{\vec B} = \frac{ \bar{\vec{r}}_\mathbf{1} } {2\bar{{r}}_\mathbf{1}^3} 
    - \bar{F}_p \frac{\bar{\vec{r}}_\mathbf{2}}{2\bar{{r}}_\mathbf{2}^3} + \xv.
    \label{eq:bfield}
\end{equation}
From here onward, we omit the bars, so that all variables are dimensionless unless stated otherwise.

\begin{figure*}[]
    \centering  
    \includegraphics[width=\textwidth]{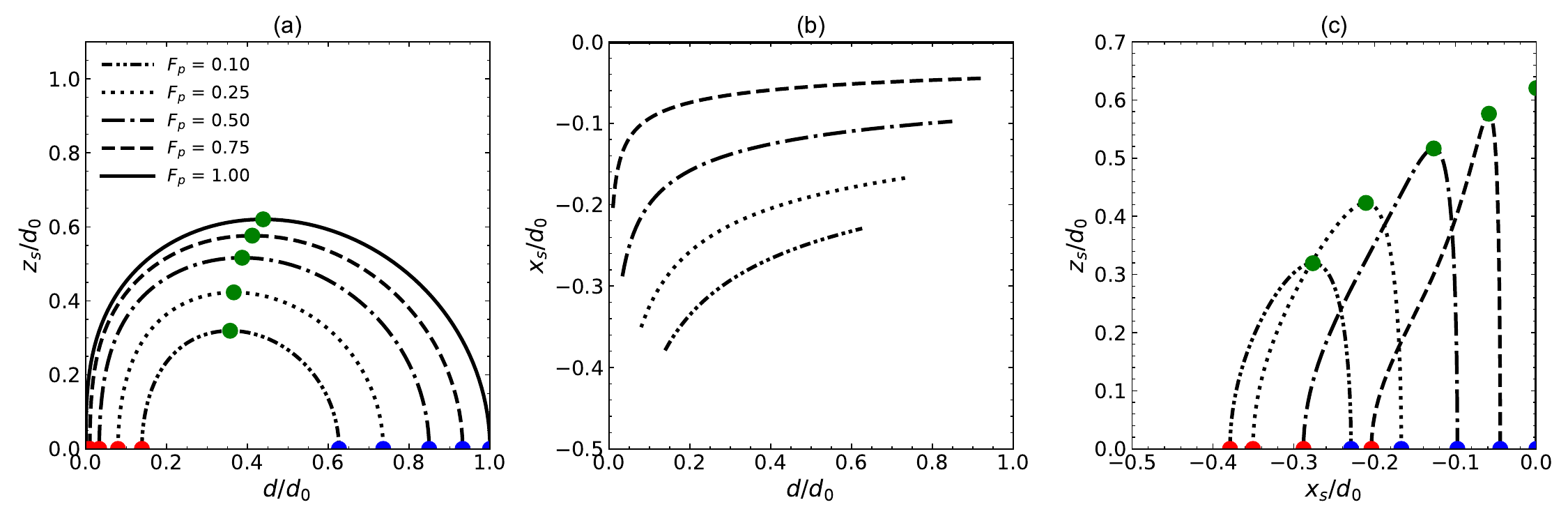}
    \caption{ 
    Position of the separator during Phase 1.
    (a) Height ($z_s$) of the separator  as a function of the half-separation ($d$) for different values of the minority flux ($F_p$). 
    The coloured points mark the locations of the maximum $z_s$ (green) and the critical points where $z_s=0$ for $d=d_{c_1}$ (blue) and $d=d_{c_2}$ (red).
    (b) Horizontal position ($x_s$) of the separator  as a function of $d$.
    (c) Positions $(x_s,z_s)$ of the separator  in the $y=0$ plane as $d$ varies for different values of $F_p$. 
    }
    \label{fig:zs}
\end{figure*}
Due to the axial symmetry about the $x$-axis, we can set $y=0$
to find the height of separator in the $xz$-plane, where  $B_x=B_z=0$, namely,
\begin{equation}
    \frac{x - d}{\rone^3} - F_p \frac{x+d}{\rtwo^3} + 2 = 0, \ \ \ \ \ 
    \frac{z}{\rone^3} - F_p \frac{z}{\rtwo^3} = 0.
    \label{eq:b=0}
\end{equation}
Solving these equations gives the  coordinates $(x_s,z_s)$ of the separator as
\begin{align}
x_s &= \frac{1}{4 {d}^{1/3}} \left( {F_p}^{2/3} - 1 \right), \ \ \ \ \ 
z_s &= \sqrt{ {d}^{2/3} - \left( \frac{{F_p}^{2/3} -1}{4{d}^{1/3}} - {d}  \right)^2}. \label{eq:xszs}
\end{align}
In Figure~\ref{fig:zs}a,b we plot $x_s$ and $z_s$ as functions of $d$ for different values of $F_p$. As $d$ decreases, the height of the separator goes from zero, to a maximum value, and then back to zero. The maximum $z_s$ can found by setting the first derivative of the above expression for $z_s$ to zero and is shown in Figure~\ref{fig:zs} as green points. It occurs for
\begin{equation}
    d_{z_{s, max}} = \left[  \frac{1}{12} \left( F_p^{2/3} + 1 + 2 \sqrt{F_p^{4/3} - F_p^{2/3} + 1} \right)  \right]^{3/4}. \label{eq:dzmax}
\end{equation}

Figure~\ref{fig:zs}a reveals that for each value of $F_p$, there are two  critical points where $z_s=0$, so that the separator has become a null point at the photosphere. To find them, we set $z_s^2=0$ and solve for $d$, which gives
\begin{equation}
    \pm d^{1/3} = \frac{F_p^{2/3} - 1}{4 d^{1/3}} - d, \nonumber
\end{equation}
where the conditions $F_p<1$ and $d>0$ rule out the positive solution, so that the above equation may be rewritten as a quadratic equation for $d^{2/3}$, namely,
\begin{equation}
  \left( d^{2/3} \right)^2 - d^{2/3} - \textstyle{\frac{1}{4}} \left( F_p^{2/3} -1 \right) = 0. \nonumber
\end{equation}
This has two solutions:
\begin{equation}
    d_{c_1} = \left[ \frac{1}{2} \left(1 + F_p^{1/3} \right) \right]^{3/2} \ \ \ \ \ \ {\rm and} \ \ \ \ \ \ d_{c_2} = \left[ \frac{1}{2} \left(1 - F_p^{1/3} \right) \right]^{3/2}.
\end{equation}
Figure~\ref{fig:zs}c shows the locations of the separator $(x_s,z_s)$ in the $y=0$ plane as the separation ($2d$) of the sources varies for different values of $F_p$. The critical points where $z_s=0$ are marked as blue for $d_{c_1}$ and red for $d_{c_2}$ points.

\begin{figure*}[h]
    \centering  
    \includegraphics[width=\textwidth]{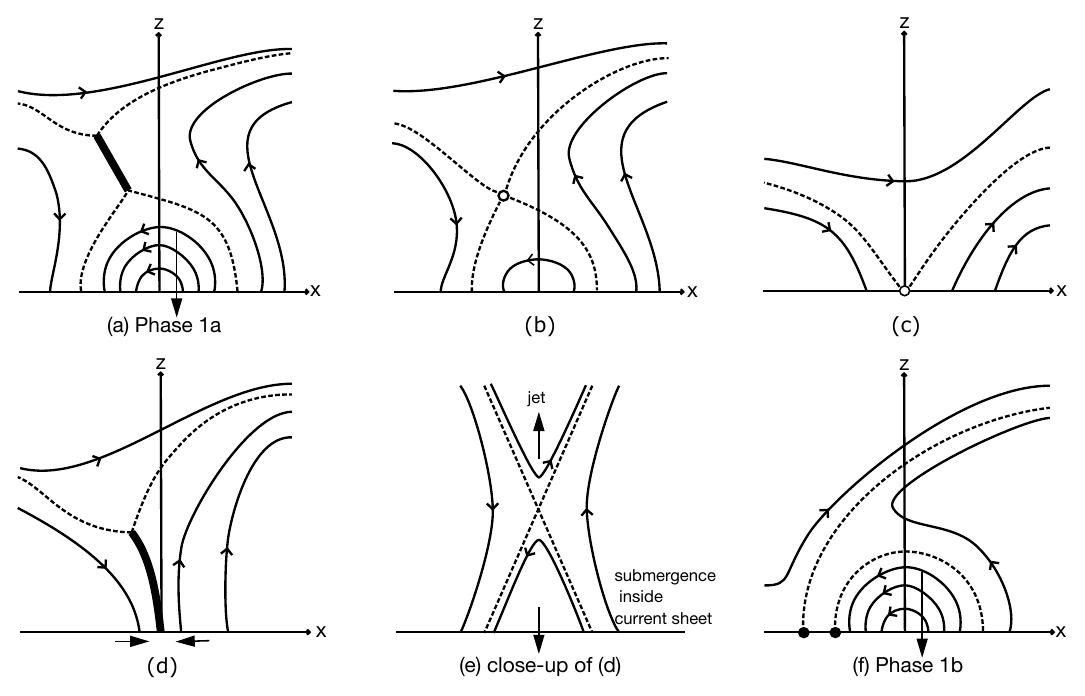}
    \caption{Simple submergence scenario for phase 2 cancellation with distributed sources (rather than point sources)  of flux:
    (a)  Magnetic field topology during phase 1a with
   submergence of the field below the separator;
    (b) Topology if reconnection ceases before submergence has been completed;
    (e) Magnetic topology if cancellation stops after large-scale submergence has finished;
    (d) Reconnection continuing when large-scale submergence has finished;
    (e) Close-up of the reconnection in the current sheet of (d), showing the production of a jet and small-scale submergence inside the current sheet;
    % (e) Magnetic topology after the cancellation of the field in (c) has finished;
    (f) Submergence during phase 1b.} 
    \label{fig:cartoonPhase2}
\end{figure*}

\subsection{Description of the Time Evolution of the System}
\label{sec:time_evolution}

\subsubsection{Phases 1a and 1b}

Having identified the location of the separator and the two critical points where the separator is located at the photosphere, we will now describe in more detail the time evolution of the system as the two  unequal and opposite flux sources approach each other within a horizontal external field. When the separation of the sources is so large that $d>d_{c_1}$ (Figure~\ref{fig:cartoon}a), there is no magnetic link between them since two separatrix surfaces separate the flux from each source into isolated domains. Two null points ($N_1$ and $N_2$) are located where these separatrices intersect the line joining the two sources in the photosphere.  

In Paper I, we identified two phases for  cancellation when the two sources have equal flux ($F_p=1$). During phase 1, a separator is formed at the photosphere and reconnection is driven at it as it rises to a maximum height and then moves back down to the photosphere, heating the plasma and accelerating a plasma jet in the process. 
During phase 2 the fluxes cancel out at the photosphere and can accelerate upwards a mixture of cool and hot plasma.
In the  $F_p=1$ case of Paper I, Figure~\ref{fig:zs}(a) shows that the separator moves back to the photosphere when the polarities fully cancel ($d=0$). 
However, in the general case where $F_p<1$, the separator returns to the photosphere at $d_{c_2}$ (red dots) before the polarities have fully cancelled. After that moment, and until the polarities fully cancel (at $d=0$), the separator bifurcates into two null points but no further reconnection and plasma heating occurs since the flux in each region remains constant.

Therefore, phase 1 displays two parts:\ phase 1a begins at $d=d_{c_1}$ (Figure~\ref{fig:cartoon}b) when a bifurcation occurs as the linear nulls  coalesce to form a second-order null, which then transforms into a separator (S).
When $d_{c_2}<d<d_{c_1}$ (Figure~\ref{fig:cartoon}c) reconnection is driven at the separator as it rises, heating plasma and accelerating jets in the process.
The two sources become magnetically more and more linked as flux is transferred into the new magnetic domain formed under the separator S. 
In addition to the rising motion, the separator moves horizontally towards the minor polarity ($-F_p$). It then passes over the minor polarity and  continues to move past it (Figure~\ref{fig:cartoon}d). 
Eventually, the separator reaches  a maximum height and  moves back downwards, reaching  the photosphere when  $d=d_{c_2}$ (Figure~\ref{fig:cartoon}e).
Until that point, reconnection occurs at the current sheet formed at the separator, locally heating  the plasma and producing various kinds of jets, similarly to what was observed by \cite*{Syntelis_etal2019} and \cite{Syntelis_Priest_2020}.

During phase 1b, the separator has moved back to the photosphere and becomes a second-order null point that bifurcates into two nulls.
For $d<d_{c_2}$ (Figure~\ref{fig:cartoon}e), while the two polarities continue to converge, the null points move across the photosphere, away from the  bifurcation location, and a new separatrix surface forms enclosing the remaining magnetic flux connecting the two polarities. The enclosed magnetic field between the two sources is not cancelled by reconnection at the null $N_2$ and no heating or plasma acceleration occurs at either $N_1$ or $N_2$ since the magnetic fluxes in each of the regions is conserved (Figure~\ref{fig:cartoon}f). 
 
This phase ends  when the two flux sources meet and the nulls coalesce.  Phase 2, however, describes the cancellation of the minority flux  with part of the majority flux source.
This cancellation  takes place either by  internal reconnection between the two cancelling polarities and  submergence of the reconnected flux or by the submergence of the field without internal reconnection. The process stops when $d=0$ and the two polarites have fully cancelled, leaving behind a separatrix surface that encloses the magnetic field of the remaining polarity of flux ($F-F_p$). For more details, see Figure~\ref{fig:cartoon}g.

Phase 2 is required to fully cancel the remaining flux under the separators to the right of $N_2$  (Figure~\ref{fig:cartoon}f). However, 
phase 2 can already begin during phase 1a, when the separator is still above the surface, or during phase 1b, when it  has sunk to the photosphere. The timing of the start of phase 2 depends on the specifics of the magnetic configuration, such as the area of the polarities, their flux content, and the distance between the two cancelling polarities since these properties dictate when the two polarities will be close enough together to  come into contact. 

As discussed in Paper I, we note that an additional possibility has been suggested by \cite{low91a}, which we call phase 0. There, after the separator appears in the solar surface in Fig. \ref{fig:cartoon}c, a current sheet grows upwards from the solar surface rather than being localised around a separator above the photosphere. It is possible that such a 'phase 0' exists for some time, if the  driving is not intermittent or if the current sheet is short enough so that it does not become fragmented due to tearing instability. We shall analyse this possibility in future.

\subsubsection{Phase 2}
\label{sec:Phase2}

There are two scenarios by which the phase 2 cancellation can occur, namely, simple submergence and  flux rope formation (a mini-filament), together with submergence. These can occur during phase 1a (Figure \ref{fig:cartoonPhase2}a) or during phase 1b (Figure \ref{fig:cartoonPhase2}f).

First, we consider the scenario of simple flux submergence (Figure~\ref{fig:cartoonPhase2}) with flux sources that are diffuse (rather than highly concentrated). One possible cause of submergence is convective flow in the photosphere with a downflow at the submergence location. Another is the   tension force of the magnetic field, provided it exceeds the magnetic pressure force and the polarities are small enough and close enough that the field lines are pulled down through the photosphere  \citep[][p. 136-141]{Parker_1979}. 

In Figure~\ref{fig:cartoonPhase2}a, we assume this process  starts during phase 1a, when reconnection is still taking place in the atmosphere. This could lead to three possible configurations. 
First of all,  cancellation could cease before the end of phase 1a,  leaving the topology shown in Figure~\ref{fig:cartoonPhase2}b. 
Secondly, it could continue until the end of phase 1a (Figure~\ref{fig:cartoonPhase2}c), leaving behind two polarities that can further cancel.
Thirdly, the field below the separator could fully submerge before reconnection at the separator is completed. Then, the lower end of the separator current sheet reaches the photosphere, separating the remaining magnetic flux of the two polarities (Figure~\ref{fig:cartoonPhase2}c). Reconnection  continues inside the current sheet, accelerating a jet upwards and submerging flux downwards (Figure~\ref{fig:cartoonPhase2}d).
Eventually, the cancellation leaves behind  magnetic  flux of magnitude $F-F_p$,  with the  topology shown in Figure~\ref{fig:cartoon}g. 
Simple submergence can also occur during phase 1b, as indicated in Figure~\ref{fig:cartoonPhase2}f. In this case, the cancellation, again, eventually leaves behind  magnetic  flux of magnitude $F-F_p$ (Figure~\ref{fig:cartoon}g).

\begin{figure*}[h]
    \centering  
    \includegraphics[width=\textwidth]{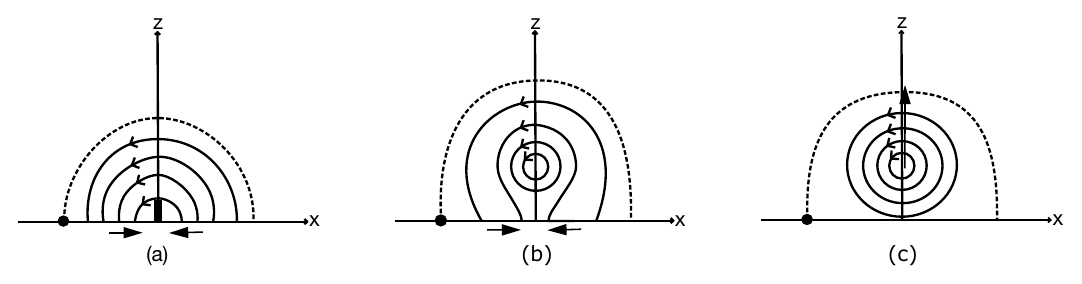}
    \caption{Flux rope scenario for phase 2 cancellation during phase 1b with distributed sources:
    (a) Magnetic field topology during phase 1b, with converging motions creating a current sheet;
    (b) Formation of flux rope containing a mini-filament and submergence of the field;
    (c) Build-up of flux rope and beginning of an eruption.} 
    \label{fig4}
\end{figure*}
\begin{figure*}[h]
    \centering  
    \includegraphics[width=0.8\textwidth]{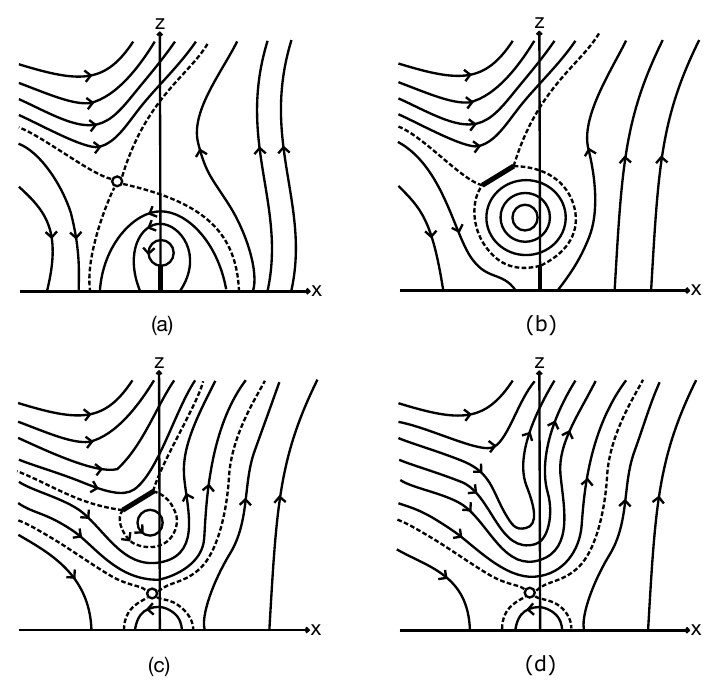}
    \caption{Flux rope scenario for phase 2 cancellation during phase 1a:
    (a) Magnetic field topology during phase 1a, with converging motions creating a current sheet;
    (b) Formation of flux rope containing a mini-filament and submergence of the field;
    (c) Reconnection of the flux rope at the overlying separator;
    (d) Topology after the flux rope erupts and diffuses.}
    \label{fig5}
\end{figure*}

The second scenario by which cancellation phase 2 can occur is through the formation of a flux rope or mini-filament, alongside the flux submergence  discussed above.
When this process starts during phase 1b (Figure~\ref{fig4}a), the new feature is the formation of a current sheet above the polarity inversion line that lies between the two polarities. Converging motions at the current sheet drive reconnection which forms a region of closed magnetic field lines in two dimensions or a magnetic island. 

If there is also a magnetic component out of the plane, that island represents a section across a twisted magnetic flux rope. Since reconnection takes place low in the atmosphere, the island or flux rope will be filled with cool plasma, which may be referred to as a  `mini-filament. As reconnection proceeds, the flux rope containing a mini-filament will rise up and expand  \citep[e.g.][]{sterling15,sterling16a,sterling16b,sterling20a}. Also, reconnection will submerge magnetic flux down through the photosphere within the current sheet and so cancel part of the initial flux (Figure~\ref{fig4}b).
As the flux rope is built up by reconnection, more and more flux will be cancelled and submerged (Figure~\ref{fig4}c). Eventually, 
 the twisted flux rope can erupt, together with jets of cool and hot plasma. The topology of such an eruption is further discussed below.
 
A flux rope can similarly be formed during phase 1a (Figure~\ref{fig5}a).
In order to erupt, the flux rope will press up against the overlying separator and form a current sheet above the flux rope, at which the flux rope will reconnect (Figure~\ref{fig5}b).  This will lead to more heating and jet production, and will allow the flux rope and the cool mini-filament to erupt along the inclined fieldlines, inducing twist by the reconnection between the flux rope with the untwisted field. Reconnection below the flux rope will assist the eruption, create an arcade field that can potentially submerge, while separating the erupting field from the non-erupting field via separatrix surfaces  (Figure~\ref{fig5}c). The erupting flux rope will fully diffuse inside the external field  (Figure~\ref{fig5}d) and the field will eventually relax. Further submergence will ultimately produce the topology in  Figure~\ref{fig:cartoon}f.

An interesting point about phase 2 cancellation concerns the role of the plasma $\beta$ (the ratio of plasma to magnetic pressure) in the later evolution of the system. As the magnetic field strength progressively decreases, the local plasma $\beta$ could become higher than unity, allowing the gas pressure forces to dominate and change the behaviour of the system. For example, a small-scale photospheric high-$\beta$ mini-filament would not necessarily erupt, but could be pushed around by the convective motions and eventually fragment or submerge by convective downdrafts.

\subsection{Energy release during phase 1}

\subsubsection{Magnetic field strength $(B_i)$ of the inflowing plasma}

At this point, we focus on phase 1a reconnection. During phase 1a, a current sheet  of length, $L,$  forms at the location of the separator, where reconnection is driven (Figure~\ref{fig:cartoon}h). 
To estimate the energy release during this process, we first need to find the input magnetic field strength at the entrance to the current sheet.
To do so, we  linearise the magnetic field and start with a two-dimensional current sheet before applying the correction that was discovered in Paper I for a three-dimensional sheet.

\begin{figure}[h]
    \centering  
    \includegraphics[width=0.8\columnwidth]{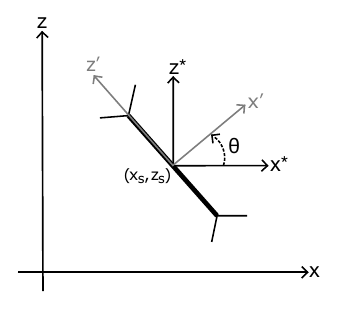}
    \caption{
        Coordinate system for the current sheet $(x^\prime, z^\prime)$ and the unrotated coordinate system  $(x^*,z^*)$ centred around ($x_s,z_s)$.
    }
    \label{fig:cartoonCS}
\end{figure}

Since the two polarities are of unequal flux, the current sheet is, unlike the case considered in paper I, generally  inclined. To simplify the analysis, we employ a coordinate system $(x^\prime, z^\prime)$ for the inclined current sheet centred around ($x_s, z_s),$ as shown in Figure~\ref{fig:cartoonCS}, where $z^\prime$ is directed along the sheet and $x^\prime$ is normal to it. Near the sheet, the magnetic field can then be written as
\begin{equation}
    B_{z^\prime} + i B_{x^\prime} = k \left( {Z^\prime}^2 + \left(\frac{L}{2} \right)^2 \right)^{1/2}
    \label{eq_2d}
,\end{equation}
in terms of the complex variable $Z^\prime = x^\prime + i z^\prime$. 
For $Z^\prime >> L$, this becomes $B_{z^\prime} + i B_{x^\prime} = k Z^\prime$, and so
\begin{equation}
\begin{bmatrix}
B_{x^\prime}  \\
B_{z^\prime}  \\
\end{bmatrix} 
= k 
\begin{bmatrix}
z^\prime  \\
x^\prime  \\
\end{bmatrix}. 
\nonumber
\end{equation}

The original coordinates are $(x,z)$, which become $(x^*,z^*)$ when translated to the centre $(x_s,z_s)$ of the current sheet, so that 
$x=x_s+x^*$ and $z=z_s+z^*$. Then the coordinates $(x^*,z^*)$ are rotated through an angle $\theta$ to give the current sheet coordinates $(x^\prime,z^\prime),$ as in Figure~\ref{fig:cartoonCS}. The relation between the magnetic field components in the two latter coordinate systems is determined by using the rotation matrix ($R_\theta$) as  
\begin{equation}
\begin{bmatrix}
B_{x^*}  \\
B_{z^*}  \\
\end{bmatrix} 
= 
R_\theta^T 
\begin{bmatrix}
B_{x^\prime}  \\
B_{z^\prime}  \\
\end{bmatrix}
=
k R_\theta^T 
\begin{bmatrix}
0 & 1  \\
1 & 0  \\
\end{bmatrix}
R_\theta
\begin{bmatrix}
x^*  \\
z^*  \\
\end{bmatrix} 
=
k 
\begin{bmatrix}
\sin 2\theta & \cos 2\theta  \\
\cos 2\theta & -\sin 2\theta  \\
\end{bmatrix}
\begin{bmatrix}
x^*  \\
z^*  \\
\end{bmatrix}. 
\nonumber
\end{equation}
By setting $a = k \sin 2 \theta$ and $b=k\cos 2 \theta$, the field components become
\begin{equation}
\begin{bmatrix}
B_{x^*}  \\
B_{z^*}  \\
\end{bmatrix} 
=
k 
\begin{bmatrix}
a & b  \\
b & -a  \\
\end{bmatrix}
\begin{bmatrix}
x^*  \\
z^*  \\
\end{bmatrix} 
\label{eq:b_system}
\end{equation}
with
\begin{equation}
    k^2 = a^2 + b^2.
    \label{eq:k}
\end{equation}
In the vicinity of the current sheet, where $x^*\approx0$ and $z^*\approx0$, the inflow magnetic field is found from Eq. (\ref{eq_2d}) to be
\begin{equation}
    B_i = \half k L = \half \sqrt{a^2+b^2} L.
    \label{eq:bi1}
\end{equation}

Next, we need to find the values of $a$ and $b$ in terms of the separatrix location ($x_s$, $z_s$) and the source half-separation ($d$). This can be accomplished by linearising either $B_x$ or $B_z$ in the vicinity of the current sheet using Eq. (\ref{eq:bfield}) and then  comparing with Eq. (\ref{eq:b_system}). The $z$-component of Eq. (\ref{eq:bfield}) is
\begin{equation}
    2B_z = \frac{z}{r_1^3} - F_p \frac{z}{r_2^3},
\end{equation}
which may be linearised around the separator by setting $x=x_s + \epsilon_1$ and $z=z_s + \epsilon_2$, where $\epsilon_1\ll x_s$ and $\epsilon_2\ll z_s$. We first note that
\begin{align}
    \frac{1}{r_{1}^3} & \approx \frac{1}{r_{1s}^3} \left( 1 - 3 \frac{x_s -d}{r_{1s}^2} \epsilon_1 - 3 \frac{z_s}{r_{1s}^2} \epsilon_2  \right), \nonumber \\
    \frac{1}{r_{2}^3} & \approx \frac{1}{r_{2s}^3} \left( 1 - 3 \frac{x_s + d}{r_{2s}^2} \epsilon_1 - 3 \frac{z_s}{r_{2s}^2} \epsilon_2  \right), \nonumber 
\end{align}
where
\begin{equation}
r_{1s} = \sqrt{(x_s-d)^2 + z_s^2} \ \ {\rm and} \ \
r_{2s} = \sqrt{(x_s+d)^2 + z_s^2}.
\label{eq:r1sr2s}
\end{equation}
Then, the linearised $B_z$ near the separator, may be written, after using Eq. (\ref{eq:b=0}), as
\begin{equation}
    B_z = \epsilon_1 \left[ -\frac{3}{2} z_s \left( \frac{x_s -d}{r_{1s}^5} - F_p \frac{x_s + d}{r_{2s}^5} \right) \right]
        + \epsilon_2 \left[ -\frac{3}{2} z_s^2 \left( \frac{1}{r_{1s}^5} - F_p \frac{1}{r_{2s}^5} \right)\right].
        \label{Bz}
\end{equation}

However, Eq. (\ref{eq:b_system}) implies that $B_z = b (x-x_s) - a (z-z_s)$, which may be compared with Eq. (\ref{Bz}) to give the values of $a$ and $b$ as
\begin{equation}
    a = \frac{3}{2} z_s^2 \left( \frac{1}{r_{1s}^5} - F_p \frac{1}{r_{2s}^5} \right) \ \ \ {\rm and} \ \ \ 
    b = - \frac{3}{2} z_s \left( \frac{x_s -d}{r_{1s}^5} - F_p \frac{x_s + d}{r_{2s}^5} \right).
    \label{eq:ab}
\end{equation}
Therefore, when the current sheet is treated as a two-dimensional structure, the magnetic field strength of the plasma flowing into the current sheet is given by Eq. (\ref{eq:bi1}) with $a$ and $b$ given by  Eqs. (\ref{eq:ab}).

Since the current sheet here is instead a three-dimensional structure, we now need to apply the correction derived in Paper I to the above two-dimensional result. The result is that the inflow magnetic field near the toroidal sheet is given, according to Eq. (\ref{eq:bi1}), by the $\half k L$ from the two-dimensional linearisation times a correction factor associated with the curvature of the current, namely,
\begin{equation}
B_i ={\half kL} (1- 0.2757\ \epsilon\log_e\epsilon),
\nonumber
\end{equation}
where $k$ is now given by Eq. (\ref{eq:k}) and $\epsilon = L/(2R_0)\ll 1$. Therefore, the magnetic field strength of the plasma flowing into the current sheet during  reconnection in phase 1a will be
\begin{equation}
B_i = {\half L \sqrt{a^2 + b^2}} (1- 0.2757\ \epsilon\log_e\epsilon),
\label{eq:bi_final}
\end{equation}
where $a$ and $b$ are given by Eqs. (\ref{eq:ab}). We note that $R_0$ is the radius of the toroidal current, and so it is the length of the line segment connecting the midpoint of the two converging polarities, $(x,z) = (0,0)$, and the separator $(x_s,z_s)$, that is,
\begin{equation}
    R_0 = \sqrt{x_s^2 + z_s^2}.
    \label{eq:r0}
\end{equation}

\subsubsection{Rate of change of flux below the separator and inflow velocity}
\label{sec:flux}

We calculate $v_{i}$ from the rate of change ($d\psi /dt$) of magnetic flux through the semicircle of radius, $z_{s}$, out of the plane of Fig.\ref{fig:cartoon}h.
This rate of change of flux becomes, after using  $\bf E+\bf v \times \bf B=\bf 0$ and Faraday's Law,
\begin{equation}
    \frac{d\psi}{dt} = -\pi z_{s}E  =  \pi z_{s}v_{i}B_{i}.
    \label{eq:dpsi_1}
\end{equation}
Then, $\psi$ may be calculated from the magnetic flux below 
$z_{S}$ through the semicircle, namely,
\begin{equation}
\psi= \int_{0}^{z_{s}}\pi z\ B_{x}(x_s,z)\ dz. \nonumber
\end{equation}
Taking its time derivative gives
\begin{align}
\frac{d\psi}{dt} &= \pi  z_s B_x(x_s, z_s) \frac{d z_s}{dt} \nonumber \\
    &+ \pi  \frac{d x_s}{dd} \int_{0}^{z_s} z \frac{\partial B_x}{\partial x_s} dz \ \ 
    +\ \  \pi  \int_{0}^{z_s} z \frac{\partial B_x}{\partial d} dz, \nonumber 
\end{align}
using that $v_0 = dd/dt=1$ is the convergence speed in dimensionless form. The first term of the above equation vanishes, since  $B_x(x_s, z_s)=0$, whereas the integrals can be calculated directly, and, following simplifications based on Eq. (\ref{eq:b=0}) at the separator, they become
\begin{align}
    & \int_{0}^{z_s} z \frac{\partial B_x}{\partial x_s} dz = 0, \ \ \ \ \ \ {\rm and}\ \ \ \ \ \ 
    \int_{0}^{z_s} z \frac{\partial B_x}{\partial d} dz = F_p\frac{z_s^2}{r_{2s}^3}, \nonumber
\end{align}
and so the rate of change of flux is
\begin{equation}
    \frac{d \psi}{dt} = \pi F_p \frac{z_s^2}{r_{2s}^3},
    \label{eq:dpsidt}
\end{equation}
where $d \psi/dt$ has been normalised with respect to $d_0^2 B_0$.
Technically, the above solution has a singular point when the separator is located above the minor polarity, but for
practical purposes, this does not cause any issues  (see  Appendix~\ref{sec:appendixA}).

Having found $d\psi/dt$, using Eq. (\ref{eq:dpsi_1}), the inflow velocity then becomes
\begin{equation}
    v_i = F_p \frac{z_s}{r_{2s}^3} \frac{2}{L}  \frac{1}{\sqrt{a^2 + b^2}}
    (1- 0.2757\ \epsilon\log_e\epsilon)^{-1}.
    \label{eq:vi_final}
\end{equation}

\subsubsection{Energy release}
\label{sec:energy}

The nature of the reconnection will dictate the length ($L$) of the current sheet  and therefore the plasma inflow speed. For example, a Sweet-Parker reconnection would result in a long diffusive region and a slow inflow of plasma. For fast reconnection, the diffusion region is much shorter and the inflow speed is larger, a fraction of the local \alfven speed \citep*[see discussion in][]{Priest_etal2018,Syntelis_etal2019}, but the current sheet includes the slow-mode shock waves, since most of the energy conversion takes place at them.

Following \citet*{Priest_etal2018}, the rate of conversion of inflowing magnetic energy into heat can be written as 
\begin{equation}
    \frac{dW}{dt}=0.8\frac{v_{i}B_{i}^{2}}{\mu}L\pi z_{s},
    \label{eq:dw_1}
\end{equation}
where $L$ is determined from Eq. (\ref{eq:vi_final}) once $v_i$ has been determined by the nature of the reconnection. For a fast reconnection, the inflow speed will be a fraction ($\alpha$, say) of the local \alfven speed, and so
\begin{equation}
    v_i = \alpha v_{Ai}, \nonumber
\end{equation}
where $v_{Ai}$ is the \alfven velocity of the inflowing plasma and $\alpha$ is typically 0.1.
In dimensional form, 
we write $v_{Ai}=v_{A0}B_{i}/B_{0}$, where $v_{A0}=B_0/\sqrt{\mu \rho_i}$. In dimensionless form, this becomes $v_{Ai}=v_{A0}B_{i}$, and so the inflow speed is
\begin{equation}
    v_i = \alpha \frac{B_i}{M_{A0}},
    \label{eq:vi}
\end{equation}
in terms of the Mach \alfven number $M_{A0} = v_0/v_{A0}$ (where $v_0=1$ in dimensionless form).
After substituting this expression into Eq. (\ref{eq:vi_final}) and using  Eq. (\ref{eq:bi_final}) for $B_i$, the length of the current sheet becomes
\begin{equation}
    L = \left(\frac{M_{A0}}{\alpha} \frac{ F_p }{ \kappa^2}\frac{z_s}{r_{2s}^3}\right)^{1/2},
    \label{eq:L}
\end{equation}
where 
\begin{equation}
    \kappa =  \frac{1}{2} \sqrt{ a^2 + b^2} 
    (1- 0.2757\ \epsilon\log_e\epsilon).
    \label{eq:kappa}
\end{equation}
Through $\epsilon,$ the expression for $L$ contains terms $L/2R_0$ on the right-hand side of the equation. So, to compute $L$, Eq. (\ref{eq:L}) can be solved numerically for $L$. Alternatively, since $L/2R_0 \ll 1$ the terms in $L/2R_0$ can be treated as corrections, so that
\begin{align}
    L &= \sqrt{ \frac{M_{A0}}{\alpha} \frac{4}{ a^2 + b^2} F_p \frac{z_s}{r_{2s}^3} } (1- 0.2757\ \epsilon\log_e\epsilon)^{-1} \nonumber \\
    & \approx L_0 (1- 0.2757\ \epsilon_0\log_e\epsilon_0)^{-1}, 
    \label{eq:L_with_L0}
\end{align}
where $\epsilon_0 = L_0 / (2 R_0)$ and $L_0 =\sqrt{4M_{A0}F_p/[\alpha(a^2 + b^2)] (z_s/r_{2s}^3)} $.

This method of calculating $L$ produces the same values as solving Eq. (\ref{eq:L}) numerically for $L$, except for  regions close to the critical points, which are discussed below. Then, after substituting for $v_i$ (Eq. (\ref{eq:vi})), $L$ (Eq. (\ref{eq:L})), and $B_i$ (Eq. (\ref{eq:bi_final})) into Eq. (\ref{eq:dw_1}),  the rate of energy conversion to heat during phase 1a finally becomes 
\begin{equation}
    \frac{dW}{dt} = 0.8  \pi \frac{M_{A0}}{\alpha} \frac{F_p^2}{ \kappa } \frac{z^3_s}{r^6_{2s}},
    \label{eq:dwdt_final}
\end{equation}
where $dW/dt$ is normalised with respect to $v_0 B_0^2 d_0^2 / \mu$.

We note that by setting $F_p=1$, all our expressions are reduced to the expressions that describe the cancellation of two equal fluxed, described in Paper I.
\begin{figure*}[h]
    \centering  
    \includegraphics[width=\textwidth]{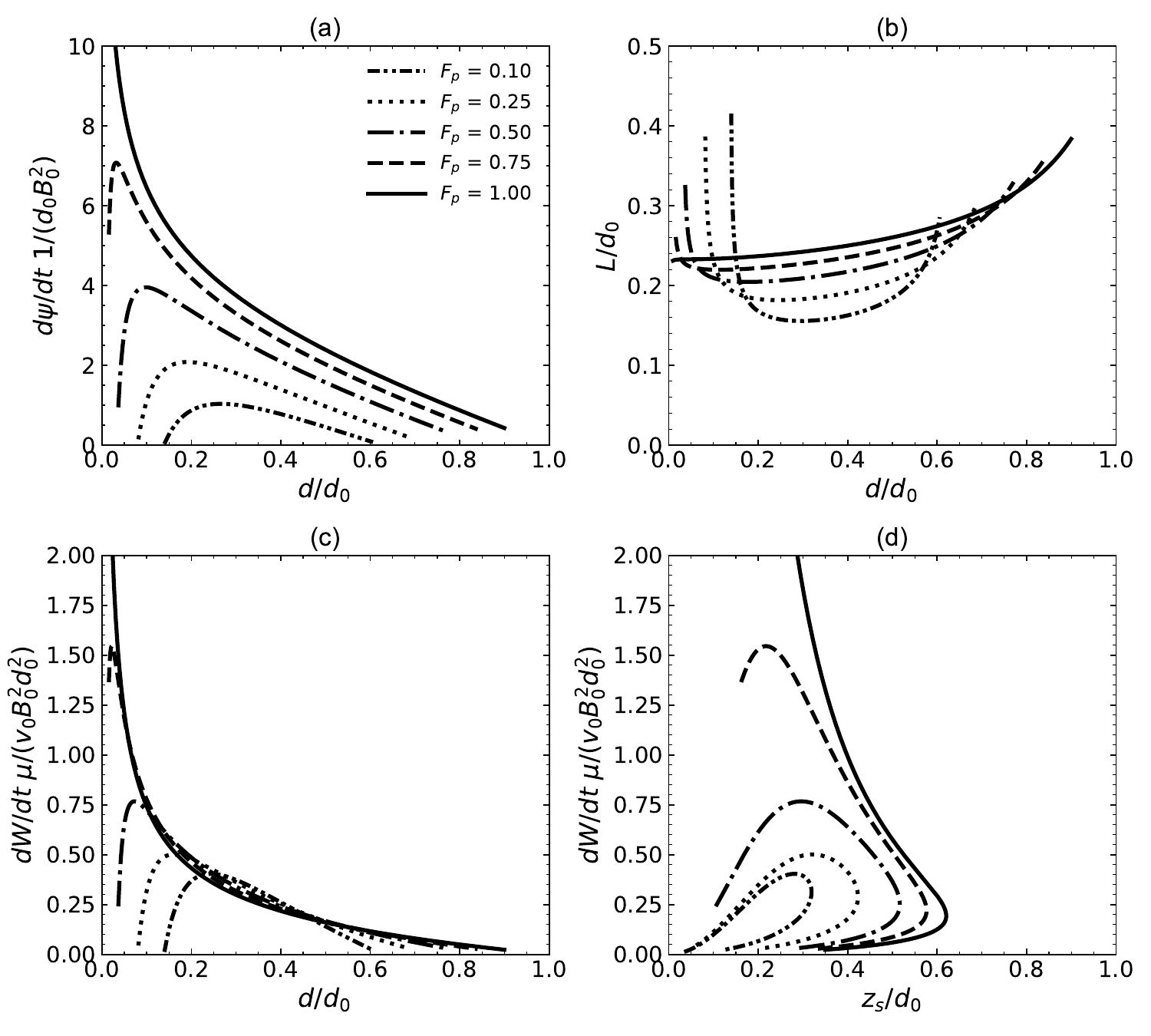}
    \caption{
    Physical parameters of cancellation for $v_0=1$, $\alpha=0.1$ and $M_{A0}=0.01$.
    (a) Rate of change ($d\psi/dt$) of flux below the separator,  for different values of the minority flux ($F_p$).
    (b) Length ($L$) of the current sheet for different values of $F_p$.
    (c) Rate ($dW/dt$) of energy conversion to heat  for different values of $F_p$. 
    (d)  $dW/dt$ for different values of $F_p$ as a function of the null height $z_s$. 
    In (b)-(d), the solution above $L_0>R_0$ is cut off.
    % For $v_0=1$, $\alpha=0.1$ and $M_{A0}=0.01$, this figure shows
    % (a) the rate of change ($d\psi^*/dt$) of flux below the separator,  for different values of the minority flux ($F_p$),
    % (b) the length ($L^*$) of the current sheet  for different values of $F_p$, and
    % (c) the rate ($dW^*/dt$) of energy conversion to heat  for different values of $F_p$. 
    % Panels (d)-(f) show the same for $M_{A0}$=0.1.
    } 
    \label{fig:psi_L_dW}
\end{figure*}

\subsubsection{Behaviour near the critical points}
\label{sec:behaviour}

The expressions for $L$, $B_i$, $d\psi / dt$ and $dW/dt$ do not always behave well near critical points ($z_s\rightarrow0$, when $d\rightarrow d_{c_1}$ or $d\rightarrow d_{c_2}$). Some of these  quantities tend to  infinity  near the critical  points  for some values of  $F_p$. This issue arises when the current sheet length becomes comparable with the distance of the separator from the photosphere and the linearisation fails. A detailed discussion on the behaviour near the critical points can be found in Appendix~\ref{sec:appendixB}.

In the numerical analysis of the analytical solutions presented in the next section, we calculate the quantities only for $L_0<R_0$, cutting off the unphysical solution around the critical points. In reality, near the critical points, each of $L$, $d\psi / dt$, $dW/dt$ should to tend to zero and produce only a small extra energy release. In fact, our $dW/dt$ solution tends to infinity only for $F_p=1$, but due to the behaviour of $L$, we cut off the solution at the linear regime for all values of $F_p$. 

% Thus, in order to make these variables vanish at the two end points where $z_s\approx0$, while preserving their maximum values in between, we replace our expressions by the starred expressions, namely,
% \begin{align}
% &\frac{d\psi^*}{dt}= \tilde{z}_s^{2\gamma}\frac{d\psi}{dt},\ \ \ 
% L^*=\tilde{z}_s^{\gamma} L,\ \ \ \frac{dW^*}{dt}=\tilde{z}_s^{4 \gamma}\frac{dW}{dt},
% \label{eq:starred}
% \end{align}
% where $\tilde{z}_s = z_s / \max(z_s)$ is the value of $z_s$ normalised with respect to its maximum value. In order for all these quantities to vanish as $d\rightarrow d_{c_1}$ or $d\rightarrow d_{c_2}$, $\gamma$ needs to be $\gamma>1$, and we choose a value for $\gamma$ of 1.1, which is found to give a reasonable shape for $L(d)$. A detailed discussion on these corrections can be found in Appendix~\ref{sec:appendixB}.

\section{Discussion}
\label{sec:discussion}

Figures~\ref{fig:psi_L_dW}(a)-(c) show the values of $d\psi/dt$, $L$, $dW/dt$ for $\alpha=0.1$ and $M_{A0}=0.01$  for different values of the fluxes ($F_p$) of the minority polarity.  The curves are cut off near $d=d_{c_1}$ and $d=d_{c_2}$, where the analysis fails since it implies unphysically that $L>R_0$.
Minority polarities with smaller flux release the energy at smaller separations, $d$, since the separator  quickly returns to the photosphere. 

These plots may be used to identify the height where the maximum $dW/dt$ occurs for different values of $F_p$. 
For the cancellation of small magnetic fragments, we can adopt an atmospheric field of, say, $B_0=10$~G, a major polarity of $F=10^{19}$~Mx and a speed of $v_0=1$~km/s, which gives  $d_0 = 5.6$ Mm, and  $v_0 B_0^2 d_0^2 / (4\pi) = 2.5\times10^{23}$~erg sec$^{-1}$. 
Therefore, for $F_p$  between $10^{18}$ and $10^{19}$~Mx, the maximum energy release rate occurs at heights of about $1.2-1.8$~Mm, that is, in the chromosphere (Figure~\ref{fig:psi_L_dW}d). 
The maximum height ($\max(z_s)$) of a separator during the reconnection is $1.8-3.5$~Mm,
while the maximum rate of energy release is $1.0-9.5\times10^{23}$~erg sec$^{-1}$.
The total energy  release during phase 1 of the cancellation may then be estimated as:
\begin{equation}
    E = \int{\frac{dW}{dt} \frac{dt}{dd} dd} = \frac{1}{v_0} \int \frac{dW}{dt} dd, \nonumber
\end{equation}
with a range of $1.6-4.2\times10^{26}$~erg. A Mach number of $M_{A0}=0.1$ instead increases by a factor of 10 the energy rates and total energies released at the same heights.
For the cancellation of smaller magnetic fragments with a majority polarity of, say, $F=10^{18}$~Mx, the flux interaction distance becomes $d_0 = 1.8$ Mm, and  $v_0 B_0^2 d_0^2 / (4\pi) = 2.5\times10^{22}$~erg sec$^{-1}$. Therefore, for $F_p$  between $10^{17}-10^{18}
$~Mx, the maximum energy release rate occurs at heights of $0.4-0.6$~Mm, namely, in the photosphere and chromosphere. 
The maximum height ($\max(z_s)$) of the separator during reconnection is $0.6-1.1$~Mm.
The maximum rate of energy release is $1-9.5\times10^{22}$~erg sec$^{-1}$, whereas the total energy released during  phase 1 cancellation is $5.2-10.3\times10^{24}$~erg. Again, $M_{A0}=0.1$ gives an order of magnitude higher energy rates and total energies.

Estimates for the energy release during phase 2 can be found in Paper I. Overall, the cancellation of small magnetic fragments produces energetic events typical of nanoflares.
We note that larger flux elements will probably consist of many finer intense flux tubes with persistent flux cancellation. In addition, during cancellation, larger flux elements can break into smaller polarities and release the energy in smaller segments. It is also important to remember that cancellation typically occurs in fits and starts. Thus, the total energy release can take place as a series of nanoflares or microflares over an extended time of hundreds or thousands of seconds, as has been reported in some observations and simulations of flux cancellation, such as \cite{Peter_etal2019} and \cite{Park_2020}. 

% The python code to calculate the quantities derived in this paper is available in GitHub\footnote{ADD URL}. \ps{[PS: Eric, do you think that its worth giving the expression derived in this paper to the public, so that someone can plugin some values and get energies etc?]}

\section{Conclusions}
\label{sec:conclusions}

Recent observations have reported that magnetic flux cancellation occurs at a significantly higher rate than previously thought. Therefore, the energy released during flux cancellation can potentially be the dominant factor in heating the atmosphere and accelerating various kinds of jets in different parts of the solar atmosphere.
The aim of the present paper has been to develop further the basic theory for energy release driven by flux cancellation.

In \citet*{Priest_etal2018}, we developed a model for energy release during the cancellation of two polarities of equal flux in the presence of a uniform horizontal magnetic field. In Paper I, we extended that theory by treating the current sheet as a series of toroidal three-dimensional current sheets, instead of  two-dimensional sheets. In this paper, we take the next step and extend the model to the case of  convergence and cancellation of two polarities of unequal flux. 

We have found that cancellation occurs in two phases. During phase 1, reconnection occurs at a separator that first moves up and then descends back down to the photosphere. We derived estimates for the energy release and the height where this occurs and found that it is released during flux cancellation in the form of nanoflares. An important new feature of this model is that when the separator moves back to the photosphere and the first phase of reconnection ceases, the two polarities have not yet fully cancelled out. Instead, the remaining magnetic flux cancels by the submergence of the field during phase 2, with reconnection occurring in or just above the photosphere between the two cancelling regions. This process can occur either via a simple reconnection and submergence of the field connecting the two polarities or by the formation and eruption of a flux rope containing a mini-filament created between the two cancelling polarities.

In our previous numerical models for the case of two cancelling polarities \citep{Syntelis_etal2019,Syntelis_Priest_2020}, we found that several types of reconnection-driven outflow can be produced during the cancellation. Depending on the height where the separator is located, reconnection can drive a wide variety of jets, whose density and temperature can vary significantly, from very cool (10,000~K) to very hot (2~MK). Similar jets would be produced during phase 1 for the general case of unequal fluxes discussed in this paper.
During phase 2, reconnection-driven jets can also form at a current sheet that extends up from the photosphere either following the simple submergence of the field of the magnetic domain connecting the two cancelling polarities or following a mini-filament eruption. In addition, the eruption of a mini-filament can form a cool eruption-driven jet, which may be more twisted than normal reconnection-driven jets.
In  the future, we would like to consider other flux geometries and use detailed numerical models to validate and advance the basic theory presented here.

%%%%%%%%%%%%%%%%%%%%%%%%%%%%%%%%%%%%%%%%%%%%%%%%%%%%%%%%
%%%%%%%%%%%%%%%%%% APPENDIX %%%%%%%%%%%%%%%%%%%%%%%%%%%%
%%%%%%%%%%%%%%%%%%%%%%%%%%%%%%%%%%%%%%%%%%%%%%%%%%%%%%%%
\begin{appendix}
\section{Rate of change of flux}
\label{sec:appendixA}
In Section \ref{sec:flux} we calculated the rate of change of  flux below the separator, namely,
\begin{equation}
\psi= \int_{0}^{z_{s}}\pi z\ B_{x}(x_s,z)\ dz. \label{psi}
\end{equation}
As the polarities converge, the separator moves towards and eventually passes over the minor polarity (Figure~\ref{fig:cartoon}b-e).
What we really need is the rate of change of the flux ($\psi_A<0$) in the domain that joins the source $F$ to the source $-F_p$. Thus, when $x_s>-d$ and the separator lies to the right of the source $-F_p$:
\begin{equation}
\psi_A = \psi \ \ \ {\rm and} \ \ \ \frac{d \psi_A}{dt} =\frac{d \psi}{dt}.
\nonumber
\end{equation}

However, when $x_s<-d$ and the separator lies to the left of the source $-F_p$, the flux calculated by Eq. ($\ref{psi}$) is:
\begin{equation}
\psi = -\psi_B, 
\nonumber
\end{equation}
in terms of the negative flux ($\psi_B$) coming in from the left to $-F_p$ (Figure~\ref{fig:cartoon}c,d). However,
\begin{equation}
\psi_A+\psi_B =-F_p,
\nonumber
\end{equation}
where $F_p>0$, and so, for $x_s<-d$, the flux we seek is
\begin{equation}
\psi_A=\psi-F_p.
\nonumber
\end{equation}
In other words, when $x_s<-d$, we need to subtract $F_p$ from the graph for $\psi$.
The result is that in both ranges for $x_s$,
\begin{equation}
\frac{d\psi_A}{dt}=\frac{d\psi}{dt}
\nonumber.
\end{equation}
When $x_s=-d$ 
and the separator lies above the minority polarity, $\psi$ is discontinuous, but the derivatives are continuous.

\begin{figure}[h]
    \centering  
    \includegraphics[width=\columnwidth]{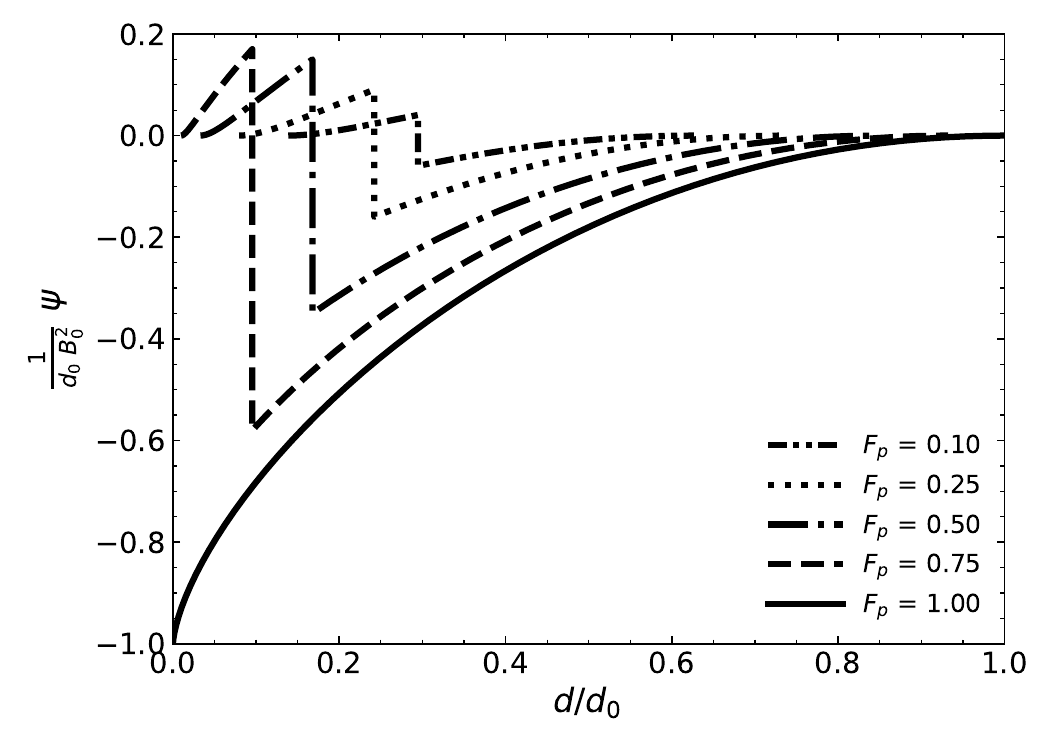}
    \caption{
        Flux ($\psi$) below the separator as a function of $d$ for different values of $F_p$, calculated from Eq. (\ref{eq:psi})
    } 
    \label{fig:appendixA}
\end{figure}

To examine this further, we calculate the flux using Eq. ($\ref{psi}$) and find its derivative. We have:
\begin{align}
\psi & = \int_{0}^{z_{s}}\pi z\ B_{x}(x_s,z)\ dz \nonumber \\
     & = \frac{\pi}{2} 
     \int_{0}^{z_{s}} z \left[  \frac{x_s - d}{ \left[ (x_s-d)^2 + z^2 \right]^{3/2}} - F_p \frac{x_s + d}{ \left[ (x_s+d)^2 + z^2 \right]^{3/2}} + 2
     \right] dz. \nonumber
\end{align}
The first two integrals are
\begin{align}
    I_1 = - \frac{x_s - d}{ \left[ (x_s-d)^2 + z^2 \right]^{1/2}} - 1 \nonumber
\end{align}
and
\begin{align}
    I_2 =  F_p \frac{x_s + d}{ \left[ (x_s+d)^2 + z^2 \right]^{1/2}} - F_p \ \mathrm{sign}(x_s+d),\nonumber
\end{align}
so the flux below the separator is
\begin{align}
    \psi = \frac{\pi}{2}  \Biggl[ &- \frac{x_s - d}{ \left[ (x_s-d)^2 + z^2 \right]^{1/2}} %\nonumber \\ 
     + F_p \frac{x_s + d}{ \left[ (x_s+d)^2 + z^2 \right]^{1/2}} \nonumber \\
     &+ z_s^2 - 1 - F_p \ \mathrm{sign}(x_s+d) \Biggl]. 
     \label{eq:psi}
\end{align}
The resulting $\psi$ is plotted in Figure~\ref{fig:appendixA} for different values of $F_p$ and clearly demonstrates the jump in value at $d=|x_s|$.
Provided $x_s\neq -d$, the derivative of the above (and therefore the value of $d \psi_A/dt$) is:
\begin{equation}
    \frac{d \psi}{dt} = \pi v_0 F_p \frac{z_s^2}{r_{2s}^3},
\end{equation}
and so our previous solution (Eq. \ref{eq:dpsidt}) is indeed valid. When $x_s= -d$, we simply define $d \psi_A/dt$ to be the limit as $x_s\rightarrow -d$ of $d \psi/dt$.

\section{Behaviour near critical points}
\label{sec:appendixB}

In this appendix, we extend the discussion in Sec.~\ref{sec:behaviour} on the behaviour of the solution near critical points. 

To do so, we examine the behaviour of $d\psi / dt$, $L$, and $dW / dt$ approaching the critical points where $z_s \rightarrow 0$ which happens as $d\rightarrow d_{c_1}$ and $d\rightarrow d_{c_2}$.

\subsection*{Case of $F_p < 1$}
\begin{figure}[h]
    \centering  
    \includegraphics[width=\columnwidth]{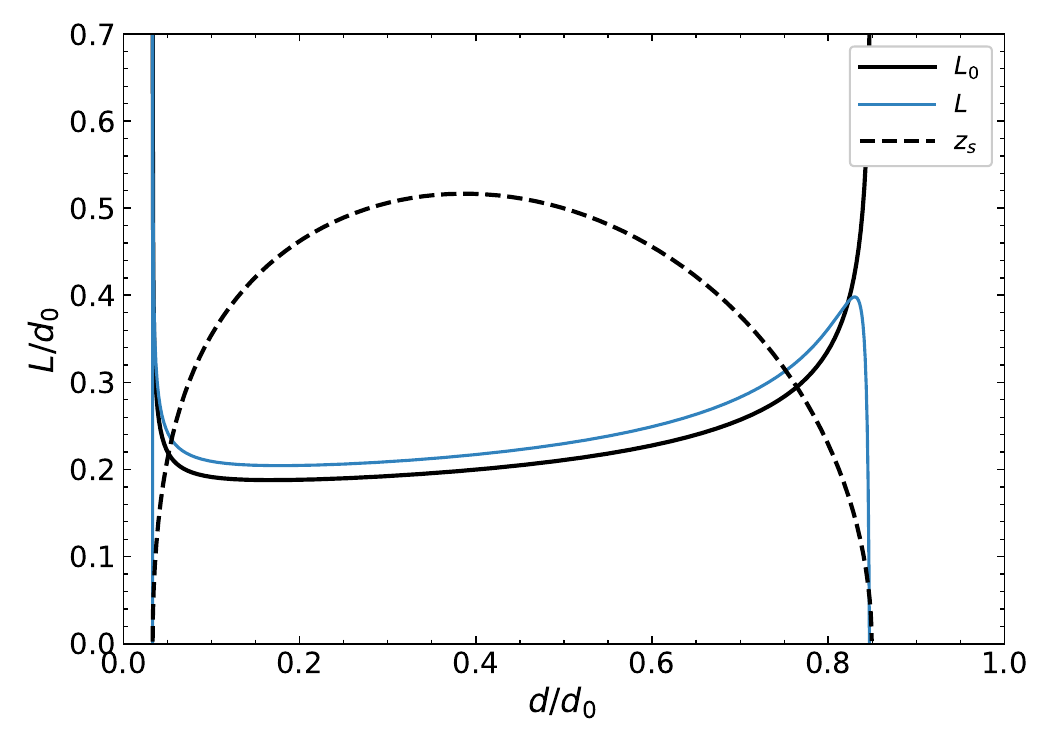}
    \caption{
        Length ($L_0$) of the current sheet without the 3D correction (solid black), the height of the separator $z_s$ (dashed black), and the length ($L$) of the current sheet with the 3D correction.
    } 
    \label{fig:appendixB}
\end{figure}

When $z_s\rightarrow0$, $x_s$ approaches a constant non-zero value (Eq~\ref{eq:xszs} and Figure~\ref{fig:zs}c),
while $r_{1s}$ and $r_{2s}$ also approach  constant non-zero values (Eq. \ref{eq:r1sr2s}). 
To examine the behaviour of $L$, we use Eq. (\ref{eq:L_with_L0}).
For $z_s\rightarrow0$, $L_0$ behaves as:
\begin{equation}
    L_0 \propto z_s^{- 1/2} \nonumber
\end{equation}
since $a\propto z_s^2$ and $b\propto z_s$. Therefore, $L_0$ will tend to infinity at the critical points. We may notice from Figure~\ref{fig:appendixB} that the linearisation fails when $z_s$ (dashed black line) becomes comparable to the $L_0$ (solid black line). After that point the separator moves to the photosphere, and $L_0$ should tend to zero instead of infinity.

The remaining terms of $L$, namely,
\begin{equation}
    \left(1-0.2757 \frac{L_0}{2 R_0}\log_e\frac{L_0}{2R_0}\right)^{-1} \nonumber
\end{equation}
tends to infinity when the denominator is zero, which is for $L_0/(2 R_0) \approx 3.13$, and eventually tends to zero when $L_0 \rightarrow\infty$ as $z_s\rightarrow0$. 
This anomalous behaviour can be seen in Figure~\ref{fig:appendixB} (blue line). We note that the singularity appears only at the lower critical point. 
% For the higher critical point, the infinity the function goes to zero faster than $\frac{L_0}{2 R_0} \approx 3.13$.

\begin{figure}[h]
    \centering  
    \includegraphics[width=\columnwidth]{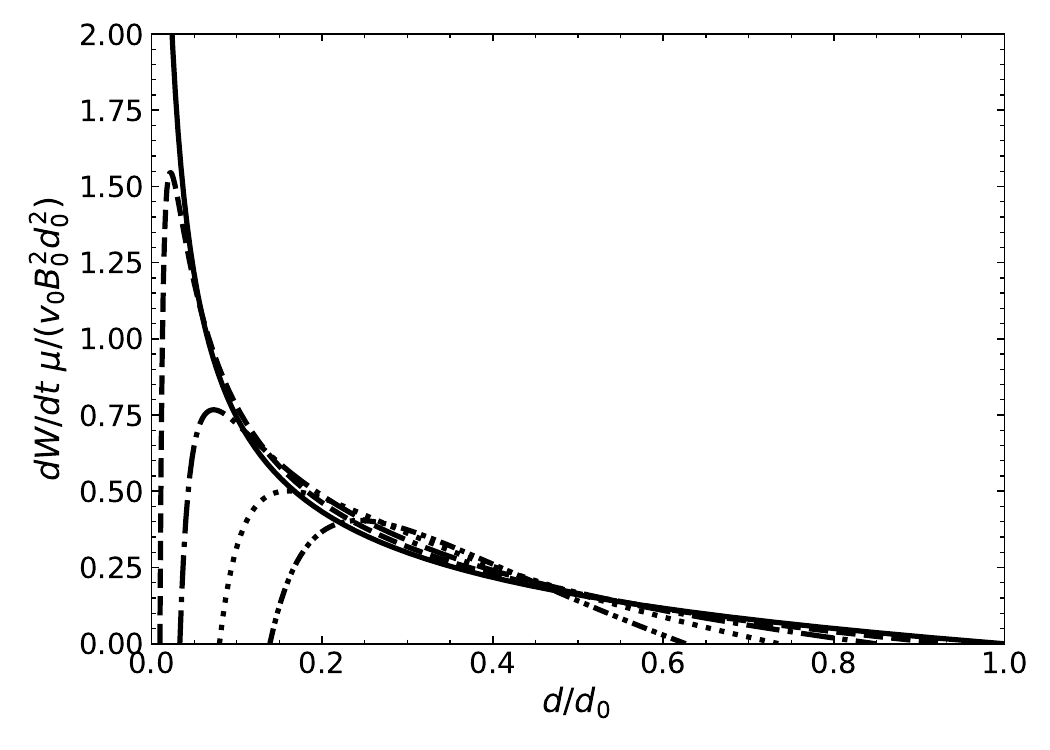}
    \caption{
        Rate ($dW/dt$) of energy conversion to heat  for different values of $F_p$, without cutting off the solution when $L_0=R_0$.
    } 
    \label{fig:appendixB2}
\end{figure}

The energy release $dW/dt$ without the 3D correction term tends to zero around the critical points for $F_p<1$. Estimating the behaviour of $dW/dt$ is more challenging with the 3D correction term, but a numerical solution of the equation shows that $dW/dt$ tends to zero before the correction term tends to infinity (Figure~\ref{fig:appendixB2}).

The rate of change of flux ${d\psi} / {dt}$ (Eq.~\ref{eq:dpsidt}) tends to zero around the critical points for $F_p<1$.

\subsection*{Case of $F_p = 1$}

When $F_p=1$,  $x_s$ is always zero and the critical points  occur for $d=0$ or $d=1$. Another important difference to the $F_p<1$ case is that $r_{1s}=r_{2s}=d^{1/3}$, which for $d=0$ tends to zero rather than a constant value. Therefore, the quantities that have terms like $1/r_{2s}$, such as  $d\psi/dt$, need to be examined with care.

As $d\rightarrow0$, $z_s \propto d^{1/3}$, the rate of change of flux behaves like $d\psi/dt \propto d^{-1/3} \propto z_s^{-1}$, and so the rate of change of flux tends to infinity for $d=0$.

The length ($L_0$)of the current sheet may now be written as:
\begin{align}
    L_0^2=\frac{4M_{A0}}{9\alpha} \frac{1}{ \sqrt{1 - d^{4/3}}},
\end{align}
which goes to infinity as $d\rightarrow 1$. Using $z_s=\sqrt{d^{2/3} - d^2} = d^{1/3} \sqrt{1 - d^{4/3}}$, we see that $L_0\propto z_s^{-1/2}$, similar to the $F_p<1$ case.

The rate of energy conversion to heat for $F_p=1$, without the 3D correction, is written as
\begin{eqnarray}
    \frac{dW}{dt}=0.8\frac{2\pi}{3} \frac{M_{A0}}{\alpha} \frac{1-d^{4/3}}{d^{2/3}},
\end{eqnarray}
which goes to infinity for $d=0$ as $dW/dt \propto d^{-2/3} \propto z_s^{-2}$.

\end{appendix}

\begin{acknowledgements}
ERP is grateful for helpful suggestions and hospitality from Pradeep Chitta, Hardi Peter, Sami Solanki and other friends in MPS G\"ottingen, where this research was initiated. P.S. acknowledge support by the ERC synergy grant ``The Whole Sun''.\\

\end{acknowledgements}

\bibliographystyle{aa}
\bibliography{bibliography}

%----------------------------------------------------------------- 
%   \begin{figure}
%   \centering
%   %%%\includegraphics[width=3cm]{empty.eps}
%       \caption{Vibrational stability equation of state
%               $S_{\mathrm{vib}}(\lg e, \lg \rho)$.
%               $>0$ means vibrational stability.
%               }
%          \label{FigVibStab}
%   \end{figure}
% %-----------------------------------------------------------------
\end{document}